\documentclass[a4paper,11pt]{article}

\usepackage{amsthm}
\usepackage{amsmath}
\usepackage{amssymb}
\usepackage{amsopn}

\newtheorem{thm}{Theorem}[section]
\newtheorem{prop}[thm]{Proposition}

\newtheorem{lem}[thm]{Lemma}
\theoremstyle{definition}

\theoremstyle{remark}

\newcommand{\lra}{\longrightarrow}
\newcommand{\ra}{\rightarrow}
\newcommand{\iny}{\hookrightarrow} 

\newcommand{\sucf}[3]{({#1})^{#3}_{{#2}=1}}

\newcommand{\N}{\mathbb{N}}

\newcommand{\R}{\mathbb{R}}
\newcommand{\C}{\mathbb{C}}

\DeclareMathOperator{\re}{Re} \DeclareMathOperator{\im}{Im}
\DeclareMathOperator{\card}{card}
\title{Deciding separability with a fixed error}
\author{David P\'{e}rez-Garc\'{\i}a \footnote{Partially supported by BMF 2001-1284}
\\
\\
{\small
Universidad Rey Juan Carlos. \'{A}rea de Matem\'{a}tica Aplicada.}\\
\small{28933 M\'{o}stoles (Madrid). Spain. E-mail address:
dperezg@escet.urjc.es}}

\date{}

\begin{document}

\maketitle

\begin{abstract}
We give a short proof of the cross norm characterization of
separability due to O. Rudolph and show how its computation, for a
fixed chosen error, can be reduced to a linear programming problem
whose dimension grows polynomially with the inverse of the error.



\end{abstract}

\section{Introduction}

Entanglement plays a key role in many of the most interesting
applications of quantum computation and quantum information
\cite{NielsenChuang}. However, there is still no procedure to
efficiently distinguish separable and entangled states. There are
two main analytical characterizations of separability: the first
one \cite{Horodeckis1} uses positive maps and the other \cite{Rud}
uses tensor norms. The problem with these characterizations is
that they are not easy to compute. This is the reason that they
are associated (by relaxing some conditions) to a number of
related \textit{computable necessary criteria of separability},
such as the PPT criterion \cite{Peres} or the CCN criterion
\cite{Rud2}, \cite{Rud3}, \cite{Rud4}. For a review see, e.g.,
\cite{Bruss}. However, up to date there is no {\it computable
characterization of separability}. In this letter we work in this
direction, by dealing with the problem of making computable the
separability characterization of \cite{Rud}. For two recent
algorithmical approaches to the separability characterization
based on positive maps (only for bipartite systems) we  refer to
\cite{DPS} and \cite{ITCE}.

To begin, let us recall some basic facts. A multipartite state can
be seen as a positive operator on a tensor product of finite
dimensional Hilbert spaces $H_1\otimes \cdots \otimes H_k$ with
trace one. We are going to call $n_j=\dim(H_j)$ and
$n=\max\{n_j\}$. Once we have fixed an orthonormal basis in each
$H_j$, we can see it as $\ell_2^{n_j,\C}$ ($\C^{n_j}$ with the
euclidean norm $\|\cdot\|_2$).  A multipartite state $\rho$ is
said to be separable if it can be prepared in a ``classical" way,
that is, if it can be written as a convex combination
$\rho=\sum_{i=1}^n \omega_i \rho^1_i\otimes\cdots\otimes
\rho^n_i$, with $\rho^j_i$ being a positive operator on $H_j$ with
trace one.

In the sequel we are going to exploit some basic facts of tensor
norms that we briefly recall here (for more information we refer
to \cite{DefantFloret93}):

If $X_1,\ldots,X_k$ are finite dimensional (real or complex)
normed spaces, by $\bigotimes_{j=1,\pi}^k X_j$ we denote the
algebraic tensor product $\bigotimes_{j=1}^k X_j$ endowed with the
projective norm
$$\pi(u):=\inf\left\{\sum_{i=1}^m \|u^1_i\|\cdots\|u^k_i\|:u=\sum_{i=1}^mu^1_i\otimes\cdots\otimes u^k_i\right\}.$$
This tensor norm is both commutative and associative, in the sense
that $\bigotimes_{j=1,\pi}^k X_j=\bigotimes_{j=1,\pi}^k
X_{\sigma(j)}$ for any permutation of the indices $\sigma$ and
that $\bigotimes_{j=1,\pi}^k \left(\bigotimes_{i_j=1,\pi}^{k_j}
X^j_{i_j}\right)=\bigotimes_{j=1,i_j=1,\pi}^{k,k_j} X^j_{i_j}$.
The projective norm $\pi$ is in duality with the injective norm
$\epsilon$, defined on $\bigotimes_{j=1}^k X_j$ as
$$\epsilon(u):=\sup\left\{\left|\sum_{i=1}^m \phi^1(u^1_i)\cdots \phi^k(u^k_i)\right|: \phi^j\in X_j^*, \|\phi_j\|\le 1\right\}$$
where $X_j^*$ denotes the topological dual of $X_j$ and
$u=\sum_{i=1}^m u^1_i\otimes \cdots\otimes u^k_i$. Moreover, this
norm is injective, in the sense that, if $Y_j$ is a subspace of
$X_j$ and $u\in \bigotimes_{j=1}^k Y_j$, we have that
$\epsilon(u)$ is the same if we consider $u$ in
$\bigotimes_{j=1,\epsilon}^k Y_j$ or in
$\bigotimes_{j=1,\epsilon}^k X_j$. Finally, $\epsilon(u)$ is just
the norm of $u$ if we see it as a $(k-1)$-linear operator
$u:X_1\times\cdots\times X_{k-1}\lra X_k^*$.

  We present now the characterization of separability given in \cite{Rud} with a simplified proof.

\begin{thm}
A multipartite state $\rho$ is separable if and only if $\rho$ is
in the closed unit ball of $\bigotimes_{j=1,\pi}^k
\mathcal{T}(H_j)$, where $\mathcal{T}(H_j)$ is the Banach space of
all trace class operators on the Hilbert space $H_j$.
\end{thm}

\begin{proof}
We only write the non-trivial part.  By definition it is clear
that the closed unit ball $B$ of $\bigotimes_{j=1,\pi}^k
\mathcal{T}(H_j)$ is the closed convex hull of
$A:=\{\rho^1\otimes\cdots\otimes \rho^k: \|\rho^j\|\le 1\}$. Since
$A$ is clearly compact, its convex hull is closed and hence
coincides with $B$. Then $\rho$, being in $B$, can be written as a
convex combination
$$\rho=\sum_{i=1}^n \omega_i\rho^1_i\otimes\cdots\otimes \rho^k_i,$$
with $\|\rho^j_i\|\le 1$. Now we reason as in \cite{Rud}:
$$1=Tr(\rho)=\sum_{i=1}^n\omega_i \prod_{j=1}^kTr(\rho^j_i)\le
\sum_{i=1}^n\omega_i\prod_{j=1}^k\|\rho^j_i\|\le 1.$$

Therefore $Tr(\rho^j_i)=\|\rho^j_i\|$ for every $i,j$, which means that $\rho^j_i$ are
positive and with trace one.
\end{proof}

With this in hand, fixing orthonormal systems in the Hilbert
spaces and using the fact that $\mathcal{T}(H_j)$ is isometric to
$H_j\otimes_\pi H_j$, deciding the separability of a density
operator is equivalent to computing the norm of the corresponding
element of
$\bigotimes_{j=1,\pi}^{k}\left(\ell_2^{n_j,\C}\otimes_\pi\ell_2^{n_j,\C}\right)=
\bigotimes_{j=1,\pi}^{2k}\ell_2^{n_j,\C}$.

The main aim of this letter then is to show how the problem of
computing a norm in $\bigotimes_{j=1,\pi}^k\ell_2^{n_j,\C}$ can be
reduced (once we have fixed the error we want to obtain) to a
\textit{linear programming problem} (LPP), which can be
efficiently solved.

We recall some terminology. By $\ell_{\infty}^{n,\R}$ we will
denote $\R^n$ with the $\sup$-norm $\|\cdot\|_{\infty}$,
$\sucf{e_i}{i}{n}$ will denote the canonical basis of $\C^n$ or
$\R^n$ and $\langle x,\phi \rangle$ or $\langle \phi,x \rangle$
will denote the duality relation $\phi(x)$, whenever $x\in X$ and
$\phi\in X^*$. Finally, we will write $\iny$ instead of simply
$\ra$ to point out that an operator is injective (and therefore
admits an inverse).

\section{Reduction to a LPP}

As a first step we treat the real case:

\begin{lem}\label{prop1}
For any $m\in \N$ and $n\in \N$, one can find (constructively) an
$N\in\N$ and a linear operator
$I:\ell_2^{n,\R}\iny\ell_{\infty}^{N,\R}$ such that $\|I\|\le 1$
and $\|I^{-1}\|\le \frac{m}{m-1}$. Moreover, we can take $N\le
(2nm+1)^n$.
\end{lem}

\begin{proof}
We take the set
    $$A:=\left\{(a_1,\ldots,a_n):a_i=\frac{h}{nm}, h=0, \pm 1,\pm 2,\ldots,
\pm nm\right\}.$$ Clearly, the cardinality of $A$ is $(2nm+1)^n$. Now, we define
$$B:=\left\{\frac{a}{\|a\|_2}: a\in A\backslash \{0\}\right\}.$$
We will see that $B$ is a $\frac{1}{m}$-covering of the unit sphere of $\ell_2^n$, that
is, for every $x\in S_{\ell_2^n}$, there exists $b\in B$ with $\|x-b\|_2\le
\frac{1}{m}$:

It is clear that, for every $x\in S_{\ell_2^n}$, there exists an
element $a\in A\backslash\{0\}$ with
$\left\|a-{x}\right\|_\infty\le \frac{1}{2nm}$. Now
    \begin{align*}
\left\|\frac{a}{\|a\|_2}-x\right\|_2=\left\|{x}-a+a-
\frac{a}{\|a\|_2}\right\|_2 \le\left\|{x}-a\right\|_2+
\left|\|a\|_2-{1}\right|.
    \end{align*}
    Using the fact that $\|\cdot\|_2\le
    \sqrt{n}\|\cdot\|_{\infty}$ we obtain that
    $$\left\|\frac{a}{\|a\|_2}-x\right\|_2\le \frac{1}{2m}+\frac{1}{2m}=\frac{1}{m}.$$

Now, it is known (see for instance \cite[page 56]{Pisier}) that
$I:\ell_2^n\iny \ell_{\infty}^{\card{B}}$, given by
$I(x)=\left(\langle x,b\rangle\right)_{b\in B}$ verifies that
$\left(1-\frac{1}{m}\right)\|x\|\le \|I(x)\|\le \|x\|$, which
means that $\|I\|\le 1$ and $\|I^{-1}\|\le \frac{m}{m-1}$.
\end{proof}

It is important to say that having $n$ in the exponent is
essential in Lemma \ref{prop1} \cite{Pisier}.

\begin{prop}\label{thm1}

If we call $c^j_{i_j}=I_j(e_{i_j})$, being
$I_j:\ell_2^{n_j,\R}\iny \ell_{\infty}^{N_j,\R}$ as in Lemma
\ref{prop1}, we have that the norm of $\rho$ in
$\bigotimes_{j=1,\pi}^k\ell_2^{n_j,\R}$ is, with a relative error
bounded by $\left(\left(\frac{m}{m-1}\right)^k-1\right)$, the
solution to the following LPP:

\noindent Maximize
$$\sum_{i_1,\ldots,i_k=1}^{n_1,\ldots,n_k}
\rho_{i_1,\ldots,i_k}\lambda_{i_1,\ldots,i_k}$$ subject to the
conditions:
$$ -1\le \sum_{i_1,\ldots,i_k=1}^{n_1,\ldots,n_k} \lambda_{i_1,\ldots,i_k}c^1_{i_1}(s_1)
\cdots c^k_{i_k}(s_k)\le 1, \quad 1\le s_j\le N_j, \quad 1\le j\le
k.$$

\end{prop}

\begin{proof}
By duality, we see the element $\rho\in \bigotimes_{j=1,\pi}^k\ell_2^{n_j,\R}$ as an
operator
$$\rho:\bigotimes_{j=1,\epsilon}^k\ell_2^{n_j,\R}\lra \R.$$ Using Lemma \ref{prop1} and
the injectivity of the $\epsilon$ norm, we have that
$I=\bigotimes_{j=1}^k
I_j:\bigotimes_{j=1,\epsilon}^k\ell_2^{n_j,\R}\iny
\bigotimes_{j=1,\epsilon}^k\ell_{\infty}^{N_j,\R}$ verifies that
$\|I\|\le 1$ and $\|I^{-1}\|\le \left(\frac{m}{m-1}\right)^k$.
Now, as $\bigotimes_{j=1,\epsilon}^k\ell_\infty^{N_j,\R}$ is
canonically isometric to $\ell_{\infty}^{N_1\cdots N_k,\R}$, we
have that the solution to the LPP gives us exactly
    $$\sup_{\|I(\lambda)\|\le 1} \langle \rho, \lambda\rangle,$$
    where $\lambda\in \bigotimes_{j=1,\epsilon}^k\ell_2^{n_j,\R}$.
        Finally, the above comments tell us that
        $$\|\rho\|=\sup_{\|\lambda\|\le 1}\langle \rho, \lambda\rangle\le
        \sup_{\|I(\lambda)\|\le 1} \langle \rho, \lambda\rangle\le \left(\frac{m}{m-1}\right)^k
        \sup_{\|\lambda\|\le 1}\langle \rho, \lambda\rangle=\left(\frac{m}{m-1}\right)^k\|\rho\|$$

\end{proof}

To see the complex case, we are going to reduce it to the real
case. The idea is the following. Again by duality, we consider the
element $\rho\in \bigotimes_{j=1,\pi}^k\ell_2^{n_j,\C}$ as an
operator
$$\rho:\bigotimes_{j=1,\epsilon}^k\ell_2^{n_j,\C}\lra \C.$$
Now,
$$\|\rho\|=\sup_{\|\lambda\|\le 1}|\langle\lambda,\rho\rangle|=
\sup_{\|\lambda\|\le 1}\re\langle\lambda,\rho\rangle=$$
$$\sup_{\|\lambda\|\le
1}\sum_{i_1,\ldots,i_k}\re(\lambda_{i_1,\ldots,i_k})\re(\rho_{i_1,\ldots,i_k})-
\sum_{i_1,\ldots,i_k}\im(\lambda_{i_1,\ldots,i_k})\im(\rho_{i_1,\ldots,i_k}).$$

Moreover, if $\lambda\in \bigotimes_{j=1,\epsilon}^k\ell_2^{n_j,\C}$, we can see it as
a $(k-1)$-linear operator
$$\lambda:\ell_2^{n_1,\C}\times\cdots\times\ell_2^{n_{k-1},\C}\lra \ell_2^{n_k,\C}.$$

If we consider the underlying real spaces we obtain that $\lambda$ can be seen also as
a $(k-1)$-linear operator
$$\tilde{\lambda}:\ell_2^{2n_1,\R}\times\cdots\times\ell_2^{2n_{k-1},\R}\lra \ell_2^{2n_k,\R},$$
with $\|\tilde{\lambda}\|=\|\lambda \|$.

The coordinates $\tilde{\lambda}_{i_1,\ldots,i_k}$ $(1\le i_j\le 2n_j)$ are as follows:

\begin{enumerate}
\item[(1)] For $1\le i_1\le n_1,\ldots,1 \le i_{k-1}\le n_{k-1}$,
$$\tilde{\lambda}_{i_1,\ldots,i_k}=\left\{\begin{array}{c}
 \re(\lambda_{i_1,\ldots,i_k}),\quad \text{if } 1\le i_k\le n_k \\
 \im(\lambda_{i_1,\ldots,i_{k-1},(i_k-n_k)}), \quad \text{if }
n_k+1\le i_k\le 2n_k \\
\end{array}\right.$$

\item[(2)] The other possibilities are completely determined by
the fact that $\lambda$ is $\C$-multilinear. For instance, if
$i_1\ge n_1+1$ and $1\le i_2\le n_2,\ldots,1\le i_k\le n_k$, we
have that
\begin{align*}
\tilde{\lambda}_{i_1,\ldots,
i_k}=\re(i\lambda_{(i_1-n_1),i_2,\dots,i_k})&=
-\im(\lambda_{(i_1-n_1),i_2,\ldots,i_k})\\&=-\tilde{\lambda}_{(i_1-n_1),i_2,\ldots,i_{k-1},(i_k+n_k)}.
\end{align*}
\end{enumerate}

Now, taking $I_j:\ell_2^{2n_j,\R}\iny  \ell_\infty^{N_j,\R}$ as in
Lemma \ref{prop1} and calling $c^j_{i_j}=I_j(e_{i_j})$ we have,
reasoning as in Proposition \ref{thm1}, that

\begin{thm}
The norm of $\rho$ in $\bigotimes_{j=1,\pi}^k\ell_2^{n_j,\C}$ is,
with a relative error bounded by
$\left(\left(\frac{m}{m-1}\right)^{k}-1\right)$, the solution of
the following LPP:

\noindent Maximize:
$$\sum_{i_1,\ldots,i_k=1}^{n_1,\ldots,n_k}
\re(\rho_{i_1,\ldots,i_k})\tilde{\lambda}_{i_1,\ldots,i_k}-
\sum_{i_1,\ldots,i_{k-1}=1}^{n_1,\ldots,n_{k-1}}\sum_{i_k=n_k+1}^{2n_k}
\im(\rho_{i_1,\ldots,i_k})\tilde{\lambda}_{i_1,\ldots,i_k}$$
subject to the conditions:
$$ -1\le \sum_{i_1,\ldots,i_k=1}^{2n_1,\ldots,2n_k} \tilde{\lambda}_{i_1,\ldots,i_k}c^1_{i_1}(s_1)\cdots
c_{i_k}(s_k)\le 1, \quad 1\le s_j\le N_j, \quad 1\le j\le k, $$
where the variables are $\tilde{\lambda}_{i_1,\ldots,i_k}$, $1\le
i_1\le n_1,\ldots,1\le i_{k-1}\le n_{k-1}, 1\le i_k\le 2n_k$, and
the other $\tilde{\lambda}_{i_1,\ldots,i_k}$ are obtained from the
variables using the conditions \emph{(2)}.
\end{thm}

\section*{Conclusion}

We have shown how the separability characterization given in
\cite{Rud} can be reduced to a linear programming problem. Though
it is a first step towards a complete computational solution to
the separability problem, it is still far from being efficient, in
the sense that, as $n$ and $k$ appear as exponents in the
dimension of the LPP, we need to assume that the number of spaces
and the dimension of them is low. That the dependence in $n$ is
exponential is not such an inconvenience in the possible
applications to quantum computing, where the dimension $n$ is
usually supposed to be $2$. Moreover, as the separability problem
has been shown to be NP-hard \cite{Gur}, this exponential
dependence in $n$ is essential to the problem and not just to our
approach. On the other hand, the exponential dependence in $k$ is
difficult to avoid, just because the dimension of the space
$\bigotimes_{j=1}^k\ell_2^n$ is $n^k$. Our approach has the
advantage that, for  fixed $n$ and $k$, the error
$\left(\left(\frac{m}{m-1}\right)^k-1\right)$ can be seen to be of
order $\frac{1}{m}$, which makes the dimension of the LPP depend
\textit{polynomially} in the inverse of the error. Moreover, our
approach works for arbitrary multipartite systems. Finally, we
think that the new techniques used here are interesting in their
own right and can lead to more efficient solutions to the
separability problem.

\providecommand{\bysame}{\leavevmode\hbox
to3em{\hrulefill}\thinspace} \providecommand{\andname}{and }

\end{document}